\documentclass[twocolappendix,appendixfloats,]{emulateapj}

\usepackage[colorlinks = true, linkcolor = blue, urlcolor  = blue, citecolor = blue, anchorcolor = blue]{hyperref}
\usepackage{hyperref}
\usepackage{apjfonts}
\usepackage{rotating}
\usepackage{amsmath}
\usepackage{color}
\usepackage{graphicx}
\usepackage[dvipsnames]{xcolor}
\usepackage{CJK}

\newcommand{\pivec}{\mbox{\boldmath $\pi$}}
\newcommand{\muvec}{\mbox{\boldmath $\mu$}}

\newcommand{\te}{t_{\rm E}}
\newcommand{\thetae}{\theta_{\rm E}}

\newcommand{\pie}{\pi_{\rm E}}

\newcommand{\dl}{D_{\rm L}}

\definecolor{brown}{rgb}{0.59, 0.29, 0.0}
\definecolor{darkgreen}{rgb}{0.0, 0.42, 0.24}
\definecolor{darkblue}{rgb}{0.01, 0.31, 0.59}
\definecolor{darkpurple}{rgb}{1.25, 0.38, 2.05}

\shorttitle{OGLE-2019-BLG-0304}
\shortauthors{Han et al.}

\begin{document}

\title{ OGLE-2019-BLG-0304: Competing Interpretations between a Planet-binary Model and 
a Binary-source + Binary-lens model}

\author{
Cheongho~Han$^{1}$, 
Andrzej~Udalski$^{2}$, 
Chung-Uk~Lee$^{3}$, 
Doeon~Kim$^{1}$,
Yoon-Hyun~Ryu$^{3}$ 
\\
(Leading authors),\\
and \\
Michael~D.~Albrow$^{4}$,   
Sun-Ju~Chung$^{3,5}$,     
Andrew~Gould$^{6,7}$,    
Kyu-Ha~Hwang$^{3}$, 
Youn~Kil~Jung$^{3}$, 
Hyoun-Woo~Kim$^{3}$,
In-Gu~Shin$^{3}$, 
Yossi~Shvartzvald$^{8}$,   
Jennifer~C.~Yee$^{9}$,     
Weicheng~Zang$^{10}$,     
Sang-Mok~Cha$^{3,11}$, 
Dong-Jin~Kim$^{3}$, 
Seung-Lee~Kim$^{3,5}$, 
Dong-Joo~Lee$^{3}$, 
Yongseok~Lee$^{3,11}$, 
Byeong-Gon~Park$^{3,5}$, 
Richard~W.~Pogge$^{7}$
\\
(The KMTNet Collaboration),\\
Przemek~Mr{\'o}z$^{2,12}$, 
Micha{\l}~K.~Szyma{\'n}ski$^{2}$,
Jan~Skowron$^{2}$,
Rados{\l}aw~Poleski$^{2}$, 
Igor~Soszy{\'n}ski$^{2}$,
Pawe{\l}~Pietrukowicz$^{2}$,
Szymon~Koz{\l}owski$^{2}$, 
Krzysztof~Ulaczyk$^{13}$,
Krzysztof~A.~Rybicki$^{2}$,
Patryk~Iwanek$^{2}$,
Marcin~Wrona$^{2}$,
Mariusz Gromadzki$^{2}$
\\
(The OGLE Collaboration)\\
}
\email{cheongho@astroph.chungbuk.ac.kr}

\affil{$^{1}$   Department of Physics, Chungbuk National University, Cheongju 28644, Republic of Korea                          }     
\affil{$^{2}$   Astronomical Observatory, University of Warsaw, Al.~Ujazdowskie 4, 00-478 Warszawa, Poland                      }     
\affil{$^{3}$   Korea Astronomy and Space Science Institute, Daejon 34055, Republic of Korea                                    }     
\affil{$^{4}$   University of Canterbury, Department of Physics and Astronomy, Private Bag 4800, Christchurch 8020, New Zealand }     
\affil{$^{5}$   Korea University of Science and Technology, 217 Gajeong-ro, Yuseong-gu, Daejeon, 34113, Republic of Korea       }     
\affil{$^{6}$   Max Planck Institute for Astronomy, K\"onigstuhl 17, D-69117 Heidelberg, Germany                                }     
\affil{$^{7}$   Department of Astronomy, The Ohio State University, 140 W. 18th Ave., Columbus, OH 43210, USA                   }     
\affil{$^{8}$   Department of Particle Physics and Astrophysics, Weizmann Institute of Science, Rehovot 76100, Israel           }     
\affil{$^{9}$   Center for Astrophysics $|$ Harvard \& Smithsonian 60 Garden St., Cambridge, MA 02138, USA                      }     
\affil{$^{10}$  Department of Astronomy and Tsinghua Centre for Astrophysics, Tsinghua University, Beijing 100084, China        }     
\affil{$^{11}$  School of Space Research, Kyung Hee University, Yongin, Kyeonggi 17104, Republic of Korea                       }     
\affil{$^{12}$  Division of Physics, Mathematics, and Astronomy, California Institute of Technology, Pasadena, CA 91125, USA    }     
\affil{$^{13}$  Department of Physics, University of Warwick, Gibbet Hill Road, Coventry, CV4 7AL, UK                           }     

\begin{abstract}
We analyze the microlensing event OGLE-2019-BLG-0304, whose light curve exhibits two distinctive 
features: a deviation in the peak region and a second bump appearing $\sim 61$~days after the main 
peak.  Although a binary-lens model can explain the overall features, it leaves subtle but 
noticeable residuals in the peak region. 
We find that the residuals can be explained by the presence of either a planetary companion located 
close to the primary of the binary lens (3L1S model) or an additional close companion to the source 
(2L2S model).  Although the 3L1S model is favored over the 2L2S model, with $\Delta\chi^2\sim 8$, 
securely resolving the degeneracy between the two models is difficult with the currently available 
photometric data.  According to the 3L1S interpretation, the lens is a planetary system, in which 
a planet with a mass $0.51^{+0.51}_{-0.23}~M_{\rm J}$ is in an S-type orbit  around a binary composed 
of stars with masses $0.27^{+0.27}_{-0.12}~M_\odot$ and $0.10^{+0.10}_{-0.04}~M_\odot$.  According to 
the 2L2S interpretation, on the other hand, the source is composed of G- and K-type giant stars, and 
the lens is composed of a low-mass M dwarf and a brown dwarf with masses $0.12^{+0.12}_{-0.05}~M_\odot$ 
and $0.045^{+0.045}_{-.019}~M_\odot$, respectively.  The event illustrates the need for through model 
testing in the interpretation of lensing events with complex features in light curves. 
\end{abstract}


\section{Introduction}\label{sec:one}

The searches for planets belonging to binary and multiple stellar systems are important because
these planets are expected to have gone through different formation and evolution processes from
those of single stars, and thus they can provide a valuable test bed to better understand the
formation and evolution processes of planets. The searches for these planets are also important 
in estimating the global frequency of planets because a majority of stars form binary or multiple 
systems, and thus the planetary frequency of multiple systems can have significant effect on the 
global planet frequency.

Planets in binary and multiple systems have been detected using various methods, including radial
velocity \citep{Correia2005}, transit \citep{Doyle2011}, pulsar-timing \citep{Thorsett1993},
eclipsing-binary \citep{Lee2009}, and microlensing \citep{Gould2014} methods. Among these
methods, microlensing provides a useful method in detecting some specific populations of planets,
especially cold planets orbiting low-mass binary stars. The microlensing sensitivity to planets in
low-mass binaries is due to the lensing property,  in which detection does not depend on the luminosity 
of host stars. Another important advantage of the microlensing method is that it enables one to detect 
planets in both S- and P-type orbits, for which a planet of the former type orbits one of the the widely 
separated binary stars, and a planet of the latter type orbits both closely spaced binary stars. The 
microlensing sensitivity to both S- and P-type planets is due to the lensing characteristics, in which 
a binary companion, regardless of the binary separation, induces a caustic in the central magnification 
region, near which a planet also induces a caustic, and thus the signatures of both the planet and 
stellar companion appear in the same peak region of high-magnification lensing light curves \citep{Lee2008, 
Han2008}.  Here caustics indicate the positions on the microlensing source plane at which the lensing 
magnification of a point source would become infinite. For this reason, high-magnification lensing events 
provide a major channel (central perturbation channel) of detecting planets in binaries.\footnote{ There 
are six discovery reports of microlensing planets in binary systems, including OGLE-2006-BLG-284 
\citep{Bennett2020}, OGLE-2008-BLG-092L \citep{Poleski2014}, OGLE-2013-BLG-0341 \citep{Gould2014}, 
OGLE-2007-BLG-349L \citep{Bennett2016}, OGLE-2016-BLG-0613L \citep{Han2017}, and OGLE-2018-BLG-1700L 
\citep{Han2020a}. Among them, four systems were detected through the central perturbation channel 
except for OGLE-2008-BLG-092L.}

Although high-magnification lensing events provide a channel to detect both S- and P-type 
planets, one is often confronted with cases in which the orbital type of the planet, S or P, 
cannot be uniquely determined. This happens because both the close- and wide-separation stellar 
companions can induce similar caustics in the central magnification region, producing companion 
signals of a similar shape. This so-called ``close--wide degeneracy'' \citep{Griest1998, Dominik1999} 
causes ambiguity not only in determining the binary separation but also in determining the star-planet 
separation. The close--wide degeneracy happened in three (OGLE-2006-BLG-284, OGLE-2016-BLG-0613L, and 
OGLE-2018-BLG-1700L) out of the total six known microlensing planetary systems in binaries, and thus 
it is a deep-seated problem in determining the orbital type of planetary systems in binaries.

The close--wide degeneracy in determining the planet and binary separations can be lifted in 
some special lens-system configurations.  The first such a case is that  the projected separation 
between the planet and host is similar to the angular Einstein radius $\thetae$ of the lens system.  
In this case, the planet induces a resonant caustic, for which the planet-host separation is uniquely 
determined.  The second case is that the source trajectory of a lensing event passes the region 
around a widely separated second stellar lens component well after (or well before) the main approach 
to the first lens component.  In this case, the source approach to the second lens component produces 
an extra bump in addition to the main peak produced by the source approach close to the first lens 
component \citep{Stefano1999}.  Then, the existence of the second bump indicates that the planet 
is in an S-type orbit.

In this work, we present the analysis of the lensing event OGLE-2019-BLG-0304.  The light curve 
of the event exhibits two distinctive features: a deviation in the peak region and a bump appearing 
long after the main peak.  A 2L1S model approximately describes the overall feature of the light 
curve, but it leaves a subtle residuals in the peak region.  If the central residual is caused by 
the perturbation of a planetary companion, then the lens is a special case in which a planet belongs 
to a binary and its orbital type is identified by the second bump.  In order to check this possibility, 
we conduct through model testing under various interpretations of the lens system and present the 
results.

\begin{figure}
\includegraphics[width=\columnwidth]{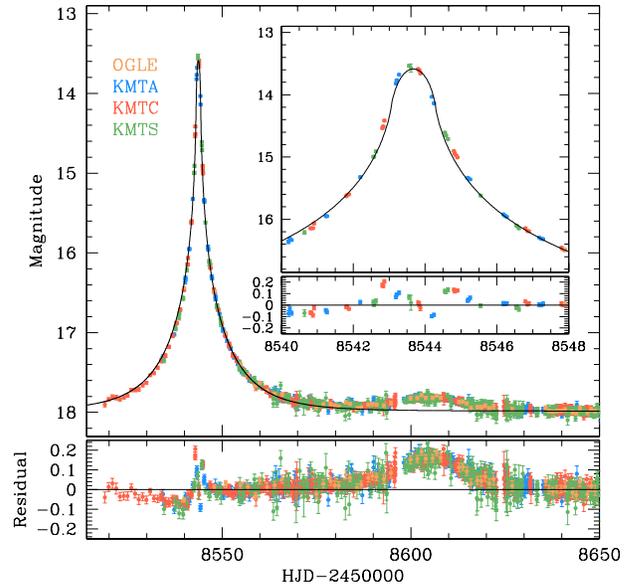}
\caption{
Light curve of the microlensing event OGLE-2019-BLG-0304. The inset shows the zoom-in view of the 
central magnification region. The curve drawn on the data points is a model obtained under a single 
lens and single source (1L1S) interpretation with the consideration of finite-source effects.  The 
lower panel shows the residuals from the 1L1S model.
}
\label{fig:one}
\end{figure}

For the presentation of the work, we organize the paper as follow.  In Sect.~\ref{sec:two}, we 
describe the observations of the lensing event and the data obtained from the observations.  We 
also mention the characteristics of the lensing event.  In Sect.~\ref{sec:three}, we describe 
various lensing models that are tested to interpret the observed lensing light curve.  The procedure 
of estimating the angular Einstein radius and relative lens-source proper motion is stated in 
Sect.~\ref{sec:four}.  In Sect.~\ref{sec:five}, we describe the procedure of estimating the physical 
lens parameters using the measured observables of the lensing event.  We discuss the reality of the 
signal in Sect.~\ref{sec:six}, and conclude in Sect.~\ref{sec:seven}.

\section{Observation and data}\label{sec:two}

The source of the event OGLE-2019-BLG-0304 is located toward the Galactic bulge field at the equatorial 
coordinates $({\rm RA}, {\rm dec})_{\rm J2000}=(17:36:06.41, -26:08:45.56)$.  The corresponding galactic 
coordinates are $(l, b)=(1^\circ\hskip-2pt.252, 3^\circ\hskip-2pt.266)$.  The brightness of the source 
had remained constant before the 2018 season with a baseline magnitude of $I_{\rm base} = 17.99$. The 
lensing-induced brightening of the source flux started during the time between the end of the 2018 
season and the beginning of the 2019 season.  The lensing light curve peaked on 2019-03-01 
(${\rm HJD}^\prime\equiv {\rm HJD}-2450000\sim 8543.5$) with a magnification of $A_{\rm peak}\sim 60$, 
and then gradually declined to the baseline. The source flux increased and peaked again on 2020-05-01 
(${\rm HJD}^\prime\sim 8605$), producing a second bump with a low magnification of $A_{\rm bump}\sim 1.2$.

\begin{figure}
\includegraphics[width=\columnwidth]{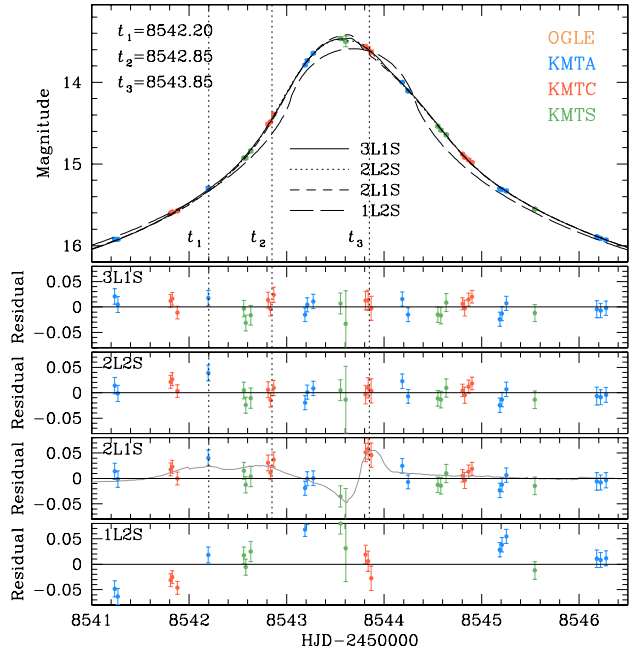}
\caption{
Four (1L2S, 2L1S, 2L2S, and 3L1S) models and their residuals in the peak region of the light curve.
The grey curve drawn in the 2L1S residual panel is the difference between the 3L1S and 2L1S models. 
The dotted vertical lines labeled by $t_1$, $t_2$, and $t_3$ indicate the three epochs of relatively 
large deviations from the 2L1S model.
}
\label{fig:two}
\end{figure}

The lensing event was first found by the Early Warning System of the Optical Gravitational Microlensing 
Experiment \citep[OGLE:][]{Udalski1994} survey on 2019-03-16.  The OGLE survey utilized the 1.3~m telescope 
located at the Las Campanas Observatory in Chile, and the telescope is equipped with a camera yielding a 
1.4~deg$^2$ field of view.  The event was rediscovered by the Korea Microlensing Telescope Network 
\citep[KMTNet:][]{Kim2016} survey in its post-season analysis \citep{Kim2018}, and it was dubbed as 
KMT-2019-BLG-2583.  The KMTNet survey employed three identical 1.6~m telescopes that are located at the 
Siding Spring Observatory (KMTA) in Australia, the Cerro Tololo Inter-American Observatory (KMTC) in 
South America, and the South African Astronomical Observatory (KMTS) in Africa.  The camera mounted on 
each KMTNet telescope yields a 4~deg$^2$ field of view.

For both surveys, observations were conducted mainly in the $I$ band, and a fraction of images were 
taken in the $V$ band for the source color measurement.  We describe the detailed procedure of the 
source color measurement in Sect.~\ref{sec:four}.  The OGLE survey of the 2019 season started on 
${\rm HJD}^\prime =8547$, 4~days before the peak, but the peak was not covered because the source 
of the event lies in a low-cadence field (BLG667), which is very infrequently observed at the beginning 
of the season due to short observing nights.  Fortunately, the peak was covered by the KMTNet survey, 
which commenced the 2019 season on ${\rm HJD}^\prime =8534$, 8518, 8534 for the KMTA, KMTC, and KMTS 
telescopes, respectively.

Neither the OGLE nor the KMT survey issued an alert of the event prior to peak, and hence no follow-up 
observations were possible, alhough this was a high-magnification event, for which the sensitivity 
to planets is high.  For example, follow-up observations of OGLE-2012-BLG-0026 \citep{Han2013} revealed 
a two-planet system, despite the fact that it peaked at a similar calendar date to OGLE-2019-BLG-0304, 
so that observations were restricted by similarly short nights. The delay of the OGLE alert was caused 
by the sparse observations of this field, as mentioned above. KMT did not issue an alert at all because 
its AlertFinder \citep{Kim2018} system began operation in 2019 on March 27, and it only searches for 
rising events. Hence, as mentioned above, the event was found in the post-season analysis.

\begin{deluxetable*}{lccccc}[thb]
\tablecaption{Lensing parameters of 1L2S, 2L1S, 2L2S, and 3L1S models\label{table:one}}
\tablewidth{480pt}
\tablehead{
\multicolumn{1}{c}{Parameter}                      &
\multicolumn{1}{c}{1L2S}                           &
\multicolumn{1}{c}{2L1S}                           &
\multicolumn{1}{c}{2L2S}                           &
\multicolumn{1}{c}{3L1S}                           
}
\startdata                                 
$\chi^2$                        &   7359.5                    &    5654.6                  &   5614.3                   &   5606.3                  \\
$t_0$ (${\rm HJD}^\prime$)      &  $8543.677 \pm 0.00266 $    &   $8543.610 \pm 0.004 $    &   $ 8543.474 \pm  0.030 $  &  $8543.599 \pm 0.005 $    \\
$u_0$ ($10^{-3}$)               &  $-0.24 \pm 0.94       $    &   $0.86 \pm 0.27      $    &   $    0.07  \pm  1.77  $  &  $0.54 \pm 0.36      $    \\
$t_{0,2}$ (${\rm HJD}^\prime$)  &  $8604.245 \pm 0.148   $    &    --                      &   $ 8543.719 \pm  0.037 $  &   --                      \\
$u_{0,2}$ ($10^{-3}$)           &  $0.831 \pm 0.026      $    &    --                      &   $    1.80  \pm  0.99  $  &   --                      \\
$\te$ (days)                    &  $14.57 \pm 0.10       $    &   $17.47 \pm 0.08     $    &   $   17.48  \pm  0.09  $  &  $17.76 \pm 0.11     $    \\
$s_2$                           &   --                        &   $3.784 \pm 0.015    $    &   $    3.785 \pm  0.016 $  &  $3.733 \pm 0.019    $    \\
$q_2$                           &   --                        &   $0.391 \pm 0.010    $    &   $    0.378 \pm  0.012 $  &  $0.363 \pm 0.012    $    \\ 
$\alpha$ (rad)                  &   --                        &   $2.954 \pm 0.002    $    &   $    2.958 \pm  0.003 $  &  $2.960 \pm 0.003    $    \\
$s_3$                           &   --                        &    --                      &                            &  $0.885 \pm 0.005    $    \\
$q_3$ ($10^{-3}$)               &   --                        &    --                      &                            &  $1.82 \pm 0.26      $    \\
$\psi$ (rad)                    &   --                        &    --                      &                            &  $2.403 \pm 0.109    $    \\
$\rho_1$ ($10^{-2}$)            &  $4.37 \pm 0.06        $    &   $1.97 \pm 0.03      $    &   $ 1.83  \pm  0.07     $  &  $1.85 \pm 0.04      $    \\   
$\rho_2$ ($10^{-2}$)            &   --                        &    --                      &   $ 2.02  \pm  0.09     $  &   --                      \\   
$q_F$                           &  $0.30 \pm 0.02        $    &    --                      &   $ 1.277 \pm  0.524    $  &   --                      
\enddata                            
\tablecomments{
${\rm HJD}^\prime\equiv {\rm HJD}-2450000$.
}
\end{deluxetable*}

\begin{figure}
\includegraphics[width=\columnwidth]{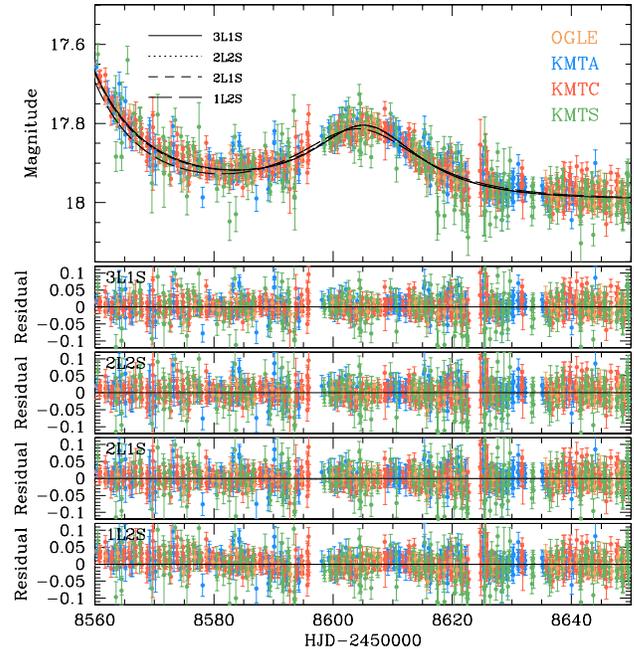}
\caption{
Four (1L2S, 2L1S, 2L2S, and 3L1S) tested models and their residuals in the region around the second 
bump of the light curve. 
}
\label{fig:three}
\end{figure}

Reduction and photometry of the data were done using the pipelines of the individual survey groups 
developed based on the difference imaging method \citep{Tomaney1996, Alard1998}: the \citet{Udalski2003} 
code using the DIA technique of \citet{Wozniak2000} for the OGLE survey and the \citet{Albrow2009} code 
for the KMTNet survey. An additional reduction was conducted for a subset of the KMTC data using the 
pyDIA code \citep{Albrow2017} to estimate the color of the source star.  Error bars of the data 
estimated by the photometry pipelines were readjusted following the routine described in \citet{Yee2012}.

Figure~\ref{fig:one} shows the light curve of OGLE-2019-BLG-0304 with the combined data obtained 
from the OGLE and KMTNet surveys.  The inset in the upper panel shows the zoom-in view of the 
peak region. Compared to the symmetric light curve of a single-lens single-source (1L1S) event, 
the light curve of the event exhibits two distinctive features. First, the light curve is 
asymmetric due to the existence of the second bump, that is centered at ${\rm HJD}^\prime\sim 8605$. 
Second, the region around the main peak, centered at ${\rm HJD}^\prime\sim 8543.5$, appears to 
exhibit deviations caused by finite-source effects, which occur when the lens passes over the 
surface of the source star.  The curve drawn over the data points is the 1L1S model obtained by 
fitting the observed data excluding the region of the second bump with the consideration of finite-source 
effects.  The model parameters are $(t_0, u_0, \te, \rho)\sim (8543.68, 0.04\times 10^{-3}, 15.3~{\rm days}, 
4.1\times 10^{-2})$, where the individual parameters indicate the peak time (in ${\rm HJD}^\prime$), 
lens-source separation at that time (normalized to $\thetae$), event timescale, and source radius (also 
normalized to $\thetae$).  Although the finite-source 1L1S model better describes the peak region than 
the point-source model, it still leaves substantial residuals, indicating that a more sophisticated model 
is needed to explain not only the second bump but also the main peak.

\section{Interpretation of anomalies}\label{sec:three}

\subsection{Three-body interpretations}\label{sec:three-one}

Considering the existence of the second bump, we first test two three-body (lens plus source) 
interpretations, in which one is that source is a binary (1L2S model) and the other is that 
the lens is a binary (2L1S model).  Besides the 1L1S parameters of $(t_0, u_0, \te, \rho)$, 
modeling the light curve under these interpretations requires one to include additional parameters. 
For the 1L2S model, these additional parameters are $(t_{0,2}, u_{0,2}, q_F, \rho_2)$, which 
represent the time of the second bump, separation between the lens and the second source at 
$t_{0,2}$, flux ratio between the two source stars ($S_1$ and $S_2$), and the normalized 
source radius of the second source, respectively \citep{Hwang2013}.  For the 2L1S model, the 
additional parameters are $(s, q, \alpha)$, which represent the separation (scaled to $\thetae$) 
and mass ratio between the two lens components ($M_1$ and $M_2$), and the angle of the source 
trajectory as measured from the $M_1$--$M_2$ axis (source trajectory angle), respectively.

\subsubsection{Binary-source (1L2S) model}

We start with the modeling under the 1L2S interpretation. In this modeling, the initial parameters
of $(t_0, u_0, \te, \rho)$ are set as those obtained from the 1L1S modeling, and the initial parameters 
of $(t_{0,2}, u_{0,2}, q_F, \rho_2)$ are set considering the time and height of the second bump in the 
light curve.  From this modeling, we find a solution that can explain the second bump. The lensing 
parameters of the 1L2S model along with the $\chi^2$ value of the fit are presented in Table~\ref{table:one}.  
We note that $\rho_2$ is not presented because the second bump does not exhibit deviations caused by 
finite-source effects.

From inspecting the residuals of the model, presented in the bottom panels 
of Figures~\ref{fig:two} (around the main peak) and \ref{fig:three} (around the second bump), it is 
found that the model still leaves substantial residuals, especially around the main peak.  This 
indicates that the 1L2S model is not a correct interpretation of the observed light curve.

\subsubsection{Binary-lens (2L1S) model}

The 2L1S modeling is carried out in two steps. In the first step, the binary parameters $s$ and
$q$ are searched for using a grid approach and the remaining parameters are searched for using
a downhill approach based on the Markov Chain Monte Carlo (MCMC) algorithm. In the second step, 
the local solutions found from the first step are refined by allowing all parameters, including
$s$ and $q$, to vary. This two-step process is needed to investigate the existence of possible
degenerate solutions.

\begin{figure}
\includegraphics[width=\columnwidth]{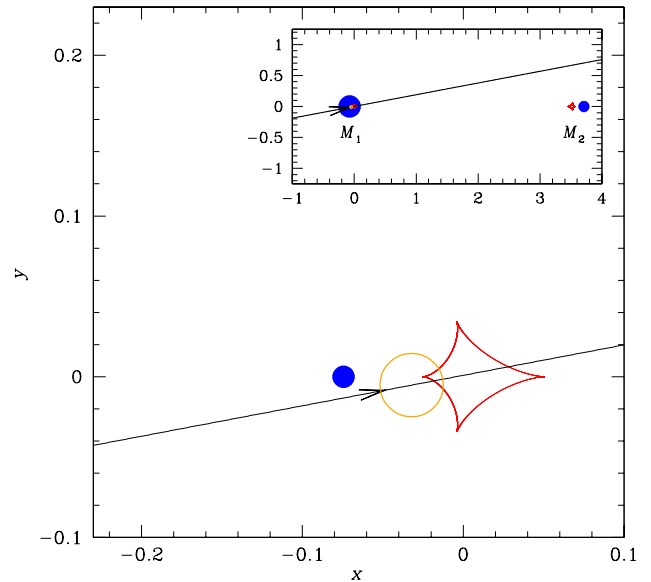}
\caption{
Configuration of the lens system for the 2L1S model.  The inset shows the whole view including 
the locations of the lens components, blue dots marked by $M_1$ and $M_2$.  The red closed figures 
represent the caustics, and the line with an arrow represents the source trajectory. The orange 
circle on the source trajectory is drawn to represent the source size relative to the caustic size. 
Lengths are scaled to the angular Einstein radius corresponding to the total mass of the lens.
\smallskip
}
\label{fig:four}
\end{figure}

From the 2L1S modeling, we find a unique model that can explain the overall features of the light 
curve by significantly reducing the residuals in both regions of the main peak and the second bump 
of the light curve. In Figures~\ref{fig:two} and \ref{fig:three}, we present the model curve and 
residuals of the 2L1S solution in the regions around the main peak and second bump, respectively. The 
best-fit lensing parameters of the solution are listed Table~\ref{table:one}, in which we mark the 
binary parameters with a subscript ``2'', i.e., $(s_2, q_2)$, to distinguish them from the parameters 
related to a possible tertiary lens component to be discussed in the following subsection. In 
Figure~\ref{fig:four}, we present  the lens system configuration of the 2L1S solution, showing the 
source trajectory with respect to the locations of the lens components and the resulting caustics.  
According to the model, the event is produced by a binary lens, in which the companion $M_2$ is 
separated in projection from the primary $M_1$ by $s\sim 3.8$, and its mass ratio to the primary is 
$q=M_2/M_1\sim 0.39$. The anomaly around the main peak of the light curve is produced by the source 
star crossing over the small caustic located close to the primary of the lens, and the second bump 
is explained by the approach of the source near the region around second caustic located close to 
the companion. We note that despite the source star crossing over the caustic during the main peak, 
usual sharp caustic-crossing features do not appear in the light curve due to the severe 
finite-source effects.  It is found that the caustic-crossing interpretation greatly reduces the 
residuals from the 1L1S model in the peak region of the light curve.  This, together with the second 
bump, strongly indicates that the lens is accompanied by a binary companion.

\begin{figure}
\includegraphics[width=\columnwidth]{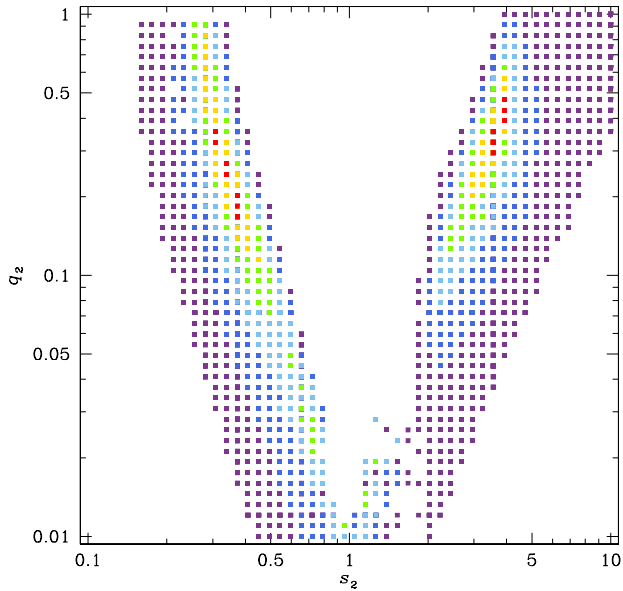}
\caption{
$\Delta\chi^2$ map in the $s_2$--$q_2$ plane obtained from the 2L1S modeling of the lensing light
curve conducted by excluding the data around the second bump.  The colors of the points represent 
those with $\leq 1n\sigma$ (red), $\leq 2n\sigma$ (yellow), $\leq 3n\sigma$ (green), $\leq 4n\sigma$ 
(cyan), $\leq 5n\sigma$ (blue), and $\leq 6n\sigma$ (purple), where $n=5$.
}
\label{fig:five}
\end{figure}

We note that the 2L1S solution would have been subject to the close--wide degeneracy, if it were not
for the data around the second bump. To check this, we conduct an additional modeling by excluding the 
data lying in the region $8570<{\rm HJD}^\prime<8635$.  Figure~\ref{fig:five} shows the $\Delta\chi^2$ 
map in the $s_2$--$q_2$ parameter plane obtained from this modeling. The map shows two distinct locals, 
in which one with $s_2>1.0$ corresponds to the wide solution presented in Table~\ref{table:one}, and 
the other with $s_2<1.0$ is the degenerate close solution. Therefore, the event is a special case, in 
which the close--wide degeneracy is clearly resolved by the existence of a second bump.

Although the 2L1S model appears to depict the overall feature of the light curve, we find that it
leaves subtle but noticeable residuals in the peak region. This can be seen in the third residual
panel (labeled as ``2L1S'') of Figure~\ref{fig:two}, which shows that the data points in this region 
deviate from the model by $\lesssim 0.05$~mag.  The three dotted vertical lines (marked by $t_1=8542.20$, 
$t_2=8542.85$, and $t_3=8543.85$) indicate the three epochs of relatively large deviations.  In order 
to check whether the deviations are caused by systematics in data, we conduct multiple rereductions 
of the data using different template images for difference imaging photometry.  We find that the 
residuals persist regardless of the reduction, suggesting that the signal is real.

We further check the possibility that the residuals from the 2L1S model around the main peak are caused
by the omission of higher-order effects in modeling. For this check, we conduct an additional modeling 
considering the microlens-parallax \citep{Gould1992} and lens-orbital \citep{Dominik1998} effects.  The 
microlens-parallax effects are caused by the positional change of an observer due to the orbital motion 
of Earth around the Sun, and the lens-orbital effects are caused by the change of the lens position due 
to the orbital motion of a binary lens. The modeling considering the microlens-parallax effect requires 
one to include two additional parameters of $\pi_{{\rm E},N}$ and $\pi_{{\rm E},E}$, which denote the 
north and east components of the microlens parallax vector $\pivec_{\rm E}=(\pi_{\rm rel}/\thetae)
(\muvec/\mu)$, respectively. Here $\pi_{\rm rel}={\rm AU}(D_{\rm L}^{-1}-D_{\rm S}^{-1})$ is the relative 
lens-source parallax, $\muvec$ denotes the relative lens-source proper motion vector, and $(D_{\rm L}, 
D_{\rm S})$ are the distances to the lens and source, respectively. Considering the lens-orbital effect 
also requires one to add two parameters of $ds/dt$ and $d\alpha/dt$, which represent the change rates of 
$s$ and $\alpha$, respectively. From this modeling, we find that the residuals from the model near the 
main peak of the light curve still persist, indicating that the cause of the residuals is not the 
higher-order effects.

\subsection{Four-body interpretations}\label{sec:three-two} 

To explain the residuals from the 2L1S model, we test two additional four-body interpretations, in 
which one is that both the lens and source are binaries (2L2S model) and the other is that the lens 
is a triple system (3L1S model).

\subsubsection{Binary source + binary lens (2L2S) model}

We conduct a 2L2S modeling considering the possibility that the discontinuous anomaly features at 
around $t_1$, $t_2$, and $t_3$ are caused by a companion to the source.  The introduction of
an additional source to a 2L1S model requires one to add extra parameters in modeling, including 
$t_{0,2}$, $u_{0,2}$, and $\rho_2$.  Here $\rho_2$ denotes the normalized source radius of the 
companion ($S_2$) to the primary source star ($S_1$).  In the modeling, we use the initial parameters 
related to $S_1$ as the ones obtained from the 2L1S modeling, because the 2L1S model describes the 
overall feature of the observed data.  The initial parameters related to $S_2$ ($t_{0,2}$, $u_{0,2}$, 
$\rho_2$, and $q_F$) are chosen considering the times and magnitudes  of the deviations from the 
2L1S model.

\begin{figure}
\includegraphics[width=\columnwidth]{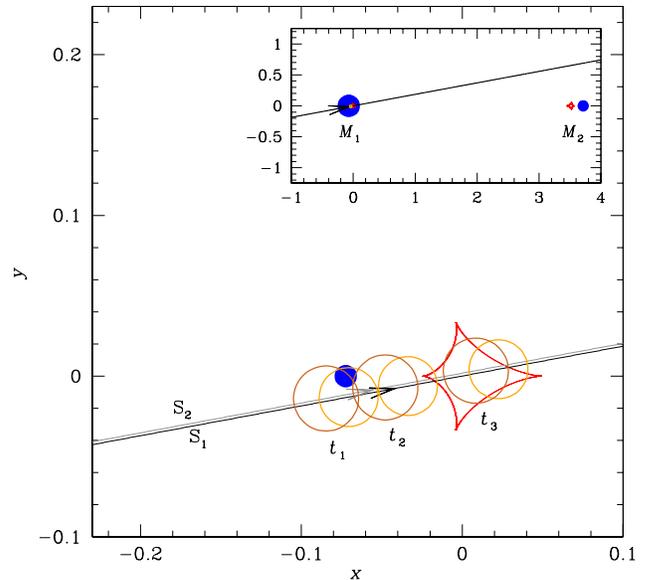}
\caption{
Configuration of the lens system for the 2L2S model.  Notations are same as those in Fig.~\ref{fig:four} 
except that there is an additional trajectory for the second source. The trajectories of the primary 
($S_1$) and companion source ($S_2$) are marked by black and grey lines, respectively.  The three 
pairs of orange and brown circles on the $S_1$ and $S_2$ trajectories represent the positions of $S_1$ 
and $S_2$ at $t_1$, $t_2$, and $t_3$, respectively. 
}
\label{fig:six}
\end{figure}

We plot the model curve of the 2L2S solution and residuals from the model in Figures~\ref{fig:two} 
and \ref{fig:three}.  The lensing parameters of the model are presented in Table~\ref{table:one}, 
and the corresponding lens system configuration is shown in Figure~\ref{fig:six}.  The configuration 
is very similar to that of the 2L1S solution, except that there is an additional trajectory for $S_2$.  
According to the 2L2S solution, $S_2$, which is brighter than $S_1$ by $\sim 28\%$ in the $I$-band flux, 
trails $S_1$  with a projected separation of $\Delta u = \{[(t_{0,2}-t_{0,1})/\te]^2 + 
(u_{0,2}-u_{0,1})^2\}=0.014$, and crosses the caustic.  We mark the positions of $S_1$ and $S_2$ 
corresponding to the times of $t_1$, $t_2$, and $t_3$ as orange and brown circles, respectively.

It is found that the model fit substantially improves with respect to the 2L1S model by introducing 
an additional source component.  First, the anomaly feature at around $t_3$, which exhibits the 
largest deviation from the 2L1S model, is explained by the caustic crossing of $S_2$.  Second, the 
anomaly feature at around $t_2$ is explained by the passage of $S_2$ through the  the positive deviation 
region extending from the left-side cusp of the caustic lying on the $M_1$--$M_2$ axis.  However, the 
model fit around the anomaly at $t_1$ remains almost unchained.  In combination, the 2L2S model results
in a better fit than the 2L1S model by $\Delta\chi^2=\chi^2_{\rm 2L1S}- \chi^2_{\rm 2L2S}=40.3$.  We 
reserve the conclusion about the model until we check an additional model to be discussed below.

\begin{figure}
\includegraphics[width=\columnwidth]{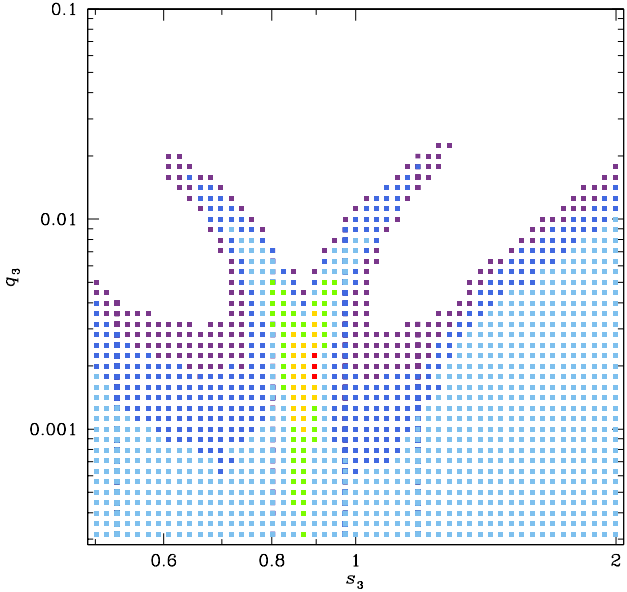}
\caption{
$\Delta\chi^2$ map in the $s_3$--$q_3$ plane obtained from the 3L1S modeling by fixing the lensing 
parameters related to $M_2$ as the parameters obtained from the 2L1S modeling.  The color coding is 
same as in Fig.~\ref{fig:six} except that $n=2$.
}
\label{fig:seven}
\end{figure}

\subsubsection{Triple-lens (3L1S) model}

We check a 3L1S interpretation because the deviations from the 2L1S model appear in the peak region,
at which an additional anomaly would occur if the lens has a tertiary component $M_3$.  The 3L1S 
modeling is carried out under the assumption that the magnification pattern of a triple lens
system can be approximated by the superposition of the patterns produced by two ($M_1$--$M_2$ and
$M_1$--$M_3$) binary pairs \citep{Bozza1999, Han2001}. Under this approximation, we search for the
parameters related to $M_3$ by fixing the lensing parameters related to $M_2$ as those of the 2L1S
model. The parameters related to $M_3$ are $(s_3, q_3, \psi)$, which represent the projected separation 
and mass ratio between $M_1$ and $M_3$, and the position angle of $M_3$ as measured from the $M_1$--$M_2$ 
axis, respectively \citep{Han2013}.

Under the superposition approximation, we first conduct thorough grid searches for $(s_3, q_3, \psi)$.
Figure~\ref{fig:seven} shows the $\Delta\chi^2$ map on the $s_3$--$q_3$ plane obtained from the grid 
search. It shows a unique and distinct local at $(s_3, q_3)\sim (0.89, 1.8\times 10^{-3})$. We refine 
the solution by allowing all parameters to vary. In Table~\ref{table:one}, we list the best-fit lensing 
parameters of the 3L1S model.  The estimated mass ratio between $M_1$ and $M_2$ of $q_2=M_2/M_1\sim 0.36$ 
is similar to the ratio estimated from the 2L1S modeling.  On the other hand, the mass ratio between 
$M_1$ and $M_3$, $q_3=M_3/M_1=(1.82\pm 0.26) \times 10^{-3}$, is very low, indicating that the tertiary 
lens component is a planetary mass object according to the 3L1S interpretation of the event.  The 
projected separation between $M_1$ and $M_3$, $s_3\sim 0.89$, is substantially smaller than the separation 
between $M_1$ and $M_2$, $s_2\sim 3.7$. This indicates that the planet is in an S-type orbit, in which 
the planet is orbiting the heavier member of a widely separated binary. From the additional modeling 
considering the higher-order effects, it is found that determining the higher-order lensing parameters 
is difficult mainly due to the short timescale, $\te \sim 18$~days, of the event.

\begin{figure}
\includegraphics[width=\columnwidth]{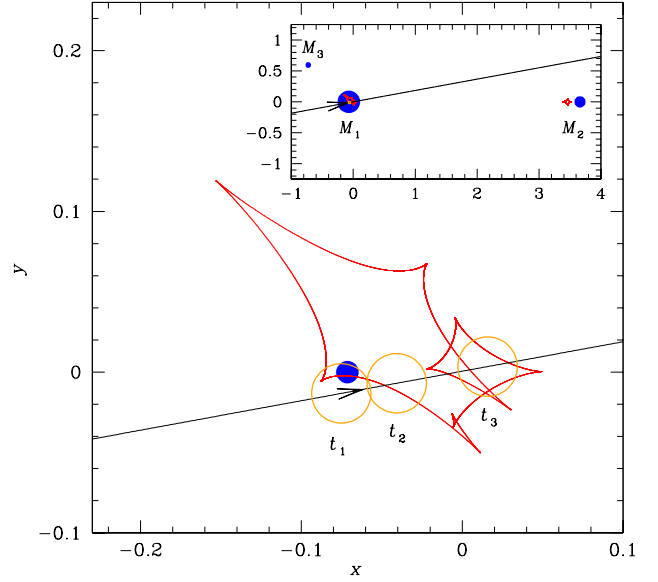}
\caption{
Lens system configuration of the 3L1S model.  The three yellow circles represent the source positions 
at $t_1$, $t_2$, and $t_3$, at which the deviations from the 2L1S model are relatively big. Other 
notations are same as those of Fig.~\ref{fig:four}.
}
\label{fig:eight}
\end{figure}

Figure~\ref{fig:eight} shows the lens system configuration of the 3L1S model. The main panel shows 
the enlarged view of the central magnification region, and the inset shows a wide view including the
positions of the individual lens components. The planet is located on the $M_1$ side with a positional
angle of $\psi\sim 138^\circ$ as measured from the $M_1$--$M_2$ axis centered at $M_1$ in a counterclockwise 
direction.  From the comparison of the 2L1S caustic shown in Figure~\ref{fig:four}, it is found that the 
tertiary lens component $M_3$ induces an additional caustic in the central region. The additional caustic 
appears to be a resonant caustic induced by a planetary companion located at a projected separation 
that is very nearly equal\footnote{ The morphology of the planetary caustic is primarily determined 
by the ratio of its separation from the host relative to the {\it host} Einstein radius 
$\theta_{\rm E,host} = \theta_{\rm E}/(1+q_2)^{1/2}$ rather than that of the total system. That is, 
$s_{3,\rm host}=s_3(1+q_2)^{1/2}=1.033$.  } to the Einstein radius.  The resulting central caustic 
appears to be the superposition of a caustics induced by a binary companion and another caustic induced 
by a planet companion, although there exist some deviations from the superposition due to the twisting 
and intersection of the caustics caused by the interference between the two caustics \citep{Gaudi1998, 
Rhie2002, Danek2015, Danek2019}.

We note that the separation of the planet from the host, $s_3$, is uniquely determined without any 
ambiguity.  In general, the central caustics induced by a pair of planetary companions with separations 
$s$ and $s^{-1}$ appear to be similar to each other due to the invariance of the binary-lens equation 
with the inversion of $s$ and $s^{-1}$.  However, this invariance breaks down as the planet separation 
is similar to $\thetae$, i.e., $s\sim 1.0$, and the lens system forms a single large resonant caustic 
\citep{Bozza1999, An2005, Chung2005}.  For OGLE-2019-BLG-0304, the planet induces such a resonant 
caustic, and thus determining $s_3$ does not suffer from the close--wide degeneracy.

\begin{figure}
\includegraphics[width=\columnwidth]{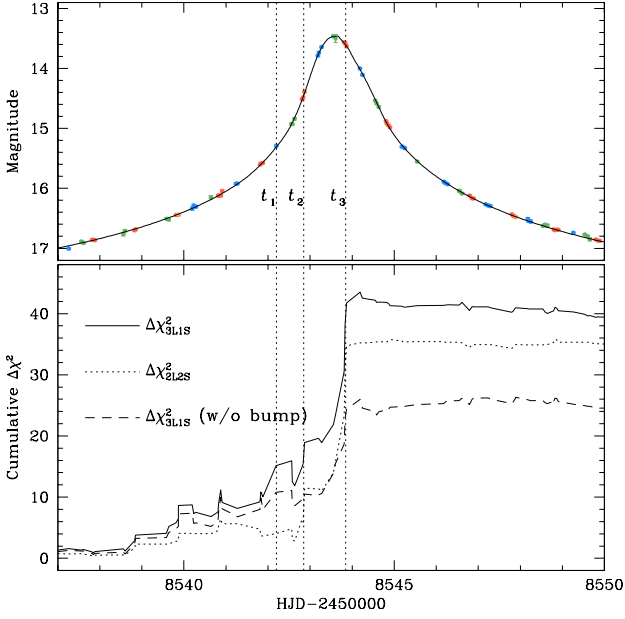}
\caption{
Cumulative distributions of $\Delta\chi_{\rm 3L1S}^2=\chi^2_{\rm 2L1S} - \chi^2_{\rm 3L1S}$
and $\Delta\chi_{\rm 2L2S}^2=\chi^2_{\rm 2L1S} - \chi^2_{\rm 2L2S}$.
The light curve in the upper panel is presented to show the region of fit improvement.  
The $\Delta\chi_{\rm 3L1S}^2$ curve drawn in dashed line is obtained from modeling with the 
exclusion of the data around the second bump.  The curve drawn over data points in the upper 
panel is the 3L1S model.
}
\label{fig:nine}
\end{figure}

It is found that the 3L1S model explains all major anomaly features.  To better show the region of 
the fit improvement around the main peak, we present the cumulative function of 
$\Delta\chi_{\rm 3L1S}^2= \chi^2_{\rm 2L1S} - \chi^2_{\rm 3L1S}$ in Figure~\ref{fig:nine}.  
For the comparison of the model fit with that of the 2L2S model, we also present the
distribution of $\Delta\chi_{\rm 2L2S}^2= \chi^2_{\rm 2L1S} - \chi^2_{\rm 2L2S}$.
The fit improvement with respect to the 2L1S model can also be seen by comparing the residuals of 
the models presented in Figure~\ref{fig:two}.  Although the fit improves throughout the rising part 
of the light curve during $8537\lesssim {\rm HJD}^\prime \lesssim 8544$, a major improvement occurs 
at the three epochs of $t_1$, $t_2$, and $t_3$.  To be noted is that the 3L1S model explains the 
anomaly at $t_1$ that could not be explained by the 2L2S model.  As a result, the fit of the 3L1S 
model is better than that of the 2L2S model by $\Delta\chi^2=8.0$.  The fit improvement over the 
2L1S model is  $\Delta\chi^2=48.3$.  We mark the source positions at the three epochs of the major 
fit improvement, i.e., $t_1$, $t_2$, and $t_3$, as orange circles in Figure~\ref{fig:eight}.  From 
this, it is found that the epoch $t_1$ corresponds to the time at which the source passes the cusp 
of the planet-induced caustic, and the other two epochs correspond to the times at which the source 
passes over the folds of the planet-induced caustic.

It is found that the second bump in the lensing light curve is important for the clear detection of the 
planetary signal.  We find this fact by comparing the fits of the two sets of 2L1S--3L1S solutions, in which 
one set of solutions are obtained from modeling with the use of all data, and the other set of solutions 
are obtained by conducting modeling  with the exclusion of the data around the region of the second bump.  
The cumulative  $\Delta\chi^2$ distribution obtained with the exclusion of the bump data is presented as a 
dashed curve (with a label ``w/o bump'' in the legend) in Figure~\ref{fig:nine}. The $\chi^2$ difference 
between the 3L1S and 2L1S models without the bump data is  $\Delta\chi^2=29.4$, which is substantially 
smaller than  $\Delta\chi^2=48.3$ when the bump data are included.

\begin{figure}
\includegraphics[width=\columnwidth]{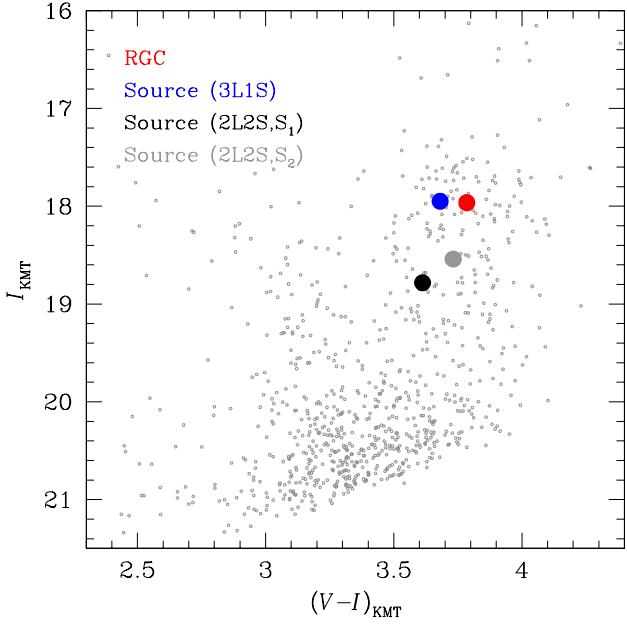}
\caption{
Locations of the source and the centroid of red giant clump (RGC) in the instrumental color-magnitude 
diagram (CMD) of stars lying in the vicinity of the source.  The CMD is constructed from the pyDIA 
photometry of the KMTC data.  The red dot is the source position estimated based on the 3L1S solution, 
while the black and grey dots are the positions of the binary source stars ($S_1$ and $S_2$) estimated 
based on the 2L2S solution.
\smallskip
}
\label{fig:ten}
\end{figure}

For the origin of the anomalies from the 2L1S model, the explanation with the existence of a low-mass 
tertiary lens component (3L1S model) is more plausible than the explanation with the existence of a 
very close companion to the source (2L2S model) for several reasons.  First, the fit of the 3L1S model 
is better than the fit of the 2L2S model.  Although $\Delta\chi^2=8.0$ is not very big, the 3L1S model 
explains all three major anomaly features (at $t_1$, $t_2$, and $t_3$), while the 2L2S model cannot 
describe one ($t_1$) of these features.  Second, the signal of a planet according to the 3L1S model 
shows up in the expected region of an event, for which the chance to detect such a signal is high, that 
is, around the peak of a high magnification event.  Third, an indirect evidence comes from the 
fact that the projected separation between $S_1$ and $S_2$ according to the 2L2S model is too close for 
the binary system to be physically stable.  The dereddened colors and magnitudes of the source stars 
are  $(V-I,I)_{0,S_1}=(0.888 \pm 0.053, 15.383 \pm 0.026)$ and $(V-I,I)_{0,S_2}=(1.007 \pm 0.049, 15.141 
\pm 0.023)$ for $S_1$ and $S_2$, respectively.  In Figure~\ref{fig:ten}, we mark the positions of $S_1$ 
and $S_2$ in the color-magnitude diagram (CMD).  These indicate that the source stars are G- and K-type 
giants with radii of $R_{S_1}\sim 6~R_\odot$ and $R_{S_2}\sim 8~R_\odot$, respectively.  On the other 
hand, the physical projected separation between the source stars is  $a_{S,\perp}=\Delta u D_{\rm S} 
\thetae\sim 4.3~R_\odot$, which is significantly smaller than $R_{S_1}+R_{S_2}\sim 14~R_\odot$.  Here 
$\Delta u = 0.014$ is the $S_1$--$S_2$ separation in units of $\thetae$, and $\thetae$ is estimated 
based on the lensing parameters of the 2L2S solution.  In principle, the source stars could avoid 
merging if $S_2$ were projected in front of or behind $S_1$.  However, unless this difference were 
considerable, the ellipsoidal distortions generated by the mutual tides of the two stars would induce 
ellipsoidal variations in the baseline light curve, which are not seen.  Fourth, in order for two source 
stars to have evolved to the same brightness on the giant branch, they would have to have nearly identical 
mass, which is also less likely.

However, the suggested reasons for the preference of the 3L1S model over the 2L2S model are either 
weak or indirect, and thus  do not collectively provide sufficient evidence to strongly favor the 3L1S 
model.  For the reason based on the $\chi^2$ difference, a $\Delta\chi^2=8.0$ is not big enough for 
strong statistical support.  This is particularly the case considering that small photometric variations 
can arise from the use of multiple data sets processed using different pipelines.  Considering that 
microlensing search algorithms are set up to systematically exclude light curves with periodically 
varying baselines, the evidence of no detectable periodic baseline variation is not strong either to 
strongly support the 3L1S model.  In order to strongly support the argument based on the low prior 
probability of observing a nearly equal-luminosity pairing of giant branch stars, one needs to know 
the relative probability of nearly equal luminosity binary giant stars versus planets in binary systems 
detectable to microlensing.  Without the information on the frequency of cool planets in binaries, this 
argument does not strongly support the 3L1S interpretation of the event.  We, therefore, consider the 
2L2S model as a viable solution.

\section{Angular Einstein radius}\label{sec:four}

The physical lens parameters can be constrained by measuring observables related to the parameters. 
The first of such observables is the event timescale, which is related to the lens parameters of the 
mass $M$ and distance $\dl$ by
\begin{equation}
\te = {\thetae\over \mu}; \ \ \ 
\thetae = \left( \kappa M \pi_{\rm rel}\right)^{1/2}; \ \ \ 
\pi_{\rm rel} = {\rm AU} \left( {1\over D_{\rm L}} - {1\over D_{\rm S}} \right).
\label{eq1}
\end{equation}
The timescale is related to the three physical parameters $M$, $\mu$ and $\dl$, and thus the constraint 
on the lens parameters is weak. The lens parameters can be more tightly constrained with the measurement 
of the angular Einstein radius, because $\thetae$ is related to two parameters of $M$ and $\dl$. In this 
section, we determine the angular Einstein radius for the use of constraining the physical lens parameters.  
In Sect.~\ref{sec:five}, we discuss the procedure of determining  $M$ and $\dl$ in detail.  We note that 
the lens parameters can be uniquely determined by additionally measuring the microlens parallax by
\begin{equation}
M= {\thetae \over \kappa\pie};\qquad
\dl = {{\rm AU}  \over  \pie\thetae + \pi_{\rm S}}.
\label{eq2}
\end{equation}
However, $\pie$ cannot be measured for OGLE-2019-BLG-0304.

The angular Einstein radius is measured from the combination of the angular source radius $\theta_*$ 
and the normalized source radius $\rho$ by
\begin{equation}
\thetae = {\theta_* \over \rho}.
\label{eq3}
\end{equation}
The value of $\rho$ is measured by analyzing the part of the lensing light curve affected by finite-source 
effects, and it is presented in Table~\ref{table:one}.  The value of $\theta_*$ is estimated from the 
color and brightness of the source star.  For the $\theta_*$ measurement, we use the standard method of 
\citet{Yoo2004}. According to this method, we first calibrate the color and magnitude of the source using 
those of the red giant clump (RGC) centroid as a reference, and then estimate $\theta_*$ using the relation 
between the color and $\theta_*$.

Figure~\ref{fig:ten} shows the locations of the source and RGC centroid in the CMD of stars lying in the 
vicinity of the source.  The source position estimated based on the 3L1S solution is marked by a red dot, 
and the positions of the individual binary source stars estimated based on the 2L2S solution are marked 
by black and grey dots.  The CMD is constructed based on the pyDIA photometry of the KMTC data set.  The 
measured instrumental colors and magnitudes of the source according to the 3L1S solution and RGC centroid 
are $(V-I, I) =(3.681\pm 0.009, 17.950\pm 0.001)$ and $(V-I, I)_{\rm RGC}=(3.785, 17.964)$, respectively.  
From the offsets in color $\Delta (V-I)$ and magnitude $\Delta I$ between the source and RGC centroid 
together with the known extinction and reddening corrected values of the RGC centroid, $(V-I, I)_{\rm RGC,0}
=(1.060, 14.564)$, toward the field from \citet{Bensby2013} and \citet{Nataf2013}, the dereddened color 
and magnitude of the source are determined as
\begin{equation}
\begin{split}
(V-I, I)_0 = & ~(V-I, I)_{\rm RGC,0} + \Delta(V-I, I)  \\
           = & ~(0.956\pm 0.009, 14.550\pm 0.001).  
\end{split}
\label{eq4}
\end{equation}
According to the 3L1S solution, the source is an RGC star with an early K spectral type.  As mentioned 
in the previous section, the stellar types of $S_1$ and $S_2$ according to the 2L2S solution are G- and 
K-type giants, respectively.

The angular source radius is deduced from the measured source color and magnitude. For this, we
first convert the measured $V-I$ color into a $V-K$ color using the color-color relation of 
\citet{Bessell1988}. We then estimate $\theta_*$ from $V-K$ using the $\theta_*$--$(V-K)$ relation 
of \citet{Kervella2004}.  The estimated source radius from this procedure is
\begin{equation}
\theta_* = 5.26 \pm 0.37~\mu{\rm as}, 
\label{eq5}
\end{equation}
for the 3L1S solution.
The angular Einstein radius estimated using the relation in Equation~(\ref{eq3}) is
\begin{equation}
\thetae = 0.285 \pm 0.020~{\rm mas}.
\label{eq6}
\end{equation}
This together with the event timescale yields the relative lens-source proper motion of
\begin{equation}
\mu =  {\thetae \over \te}=  5.86 \pm 0.41~{\rm mas}~{\rm yr}^{-1}.
\label{eq7}
\end{equation}

The Einstein radius and  proper motion based on the 2L2S solution result in different values from 
those estimated based on the 3L1S solution.  This is because the flux from the source is divided into 
two roughly equal-luminosity stars, and thus the source radii of the individual source stars are smaller 
than that of a single source for the 3L1S solution.  The estimated angular radius of $S_1$, Einstein 
radius, and proper motion are $\theta_{*,S_1}=3.24\pm 0.28$, $\thetae=\theta_{*,S_1}/\rho_1=0.177\pm 
0.016$~mas, and $\mu=3.69\pm 0.32~{\rm mas}~{\rm yr}^{-1}$, respectively.  We note that the values of 
$\thetae$ and $\mu$ are smaller than those estimated from the 3L1S solution due to the smaller source 
radius.

\section{Physical lens parameters}\label{sec:five}

Although it is difficult to uniquely determine $M$ and $\dl$ due to the difficulty of measuring $\pie$, 
the lens parameters can still be constrained with the measured observables of $\te$ and $\thetae$. For 
this, we conduct a Bayesian analysis of the event based on the priors of the lens mass function and 
Galactic model of the physical and dynamic distributions.

\begin{figure}[t]
\includegraphics[width=\columnwidth]{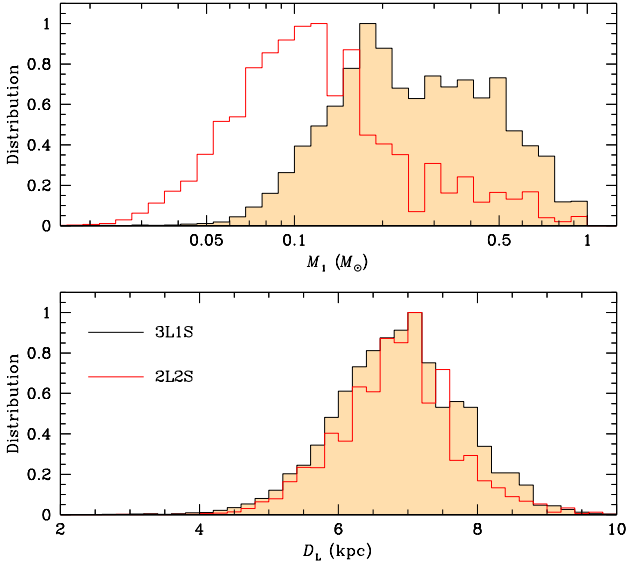}
\caption{
Bayesian posteriors of the primary lens mass $M_1$ (upper panel) and the distance to the lens $\dl$ 
(lower panel).  In each panel, the shaded black curve is the distribution based on the 3L1S solution, 
while the unshaded red curve is the distribution based on the 2L2S solution.  
\smallskip\smallskip
}
\label{fig:eleven}
\end{figure}

In the Bayesian analysis, we conduct a Monte Carlo simulation to produce a large number of artificial 
lensing events. The mass function used in the simulation is adopted from \citet{Zhang2020} for stellar 
and brown-dwarf lenses and \citet{Gould2000} for remnant lenses, including  white dwarfs, neutron stars, 
and black holes.  The lens objects are physically distributed based on the modified \citet{Han2003} 
model, in which the distribution of disk matter is modified from the original version using the model 
of \citet{Bennett2014}.  The motion of the lens is assigned using the dynamical model of \citet{Han1995}. 
The proper motion of the source, $\muvec_{\rm S} ({\rm RA,dec})= (-6.737\pm 1.721, -9.277\pm 1.284) 
~{\rm mas}~{\rm yr}^{-1}$, is known from the measurement by {\it Gaia} \citep{Gaia2018}, and thus we 
consider the measured source proper motion in the computation of the relative lens-source motion.  
With these priors, we produce $4\times 10^7$ artificial events, and then construct probability 
distributions of $M$ and $\dl$ for events with $\te$ and $\thetae$ values lying within the ranges of 
the measured observables. Then, the median values of the probability distributions are presented as 
representative values of the physical lens parameters, and the uncertainties are estimated as the 
$1\sigma$ range of the probability distributions, that is, 16\% for the lower limit and 84\% for the 
upper limit.  Considering that the 3L1S and 2L2S solutions result in similar fits to the observed data, 
we conduct two sets of analysis based on the two solutions.

The shaded curves in Figure~\ref{fig:eleven} shows the posterior distributions of the $M_1$ (upper panel) 
and $\dl$ (lower panel) based on the 3L1S solution.  The estimated masses of the individual lens components 
are
\begin{equation}
M_1=  0.27^{+0.27}_{-0.12}~M_\odot,\ \
M_2=  0.10^{+0.10}_{-0.04}~M_\odot,\ \ 
M_3=  0.51^{+0.51}_{-0.23}~M_{\rm J}, 
\label{eq8}
\end{equation}
indicating that the lens is a planetary system in which a giant plant belongs to a binary composed
of a mid M dwarf and a late M dwarf.  The estimated distance to the lens is
\begin{equation}
\dl = 6.98^{+0.91}_{-0.90}~{\rm kpc}.
\label{eq9}
\end{equation}
The projected physical separations of the stellar ($M_2$) and planetary ($M_3$) companions from the 
primary ($M_1$) are
\begin{equation}
a_{\perp,2} = s_2\dl \thetae = 5.19^{+5.87}_{-4.52}~{\rm AU},\  
a_{\perp,3} = s_3\dl \thetae = 1.23^{+1.39}_{-1.07}~{\rm AU}, 
\label{eq10}
\end{equation}
respectively. Considering that $a_{\perp,3}$ is substantially smaller than $a_{\perp,2}$, the planet is 
very likely to be in an S-type orbit around the heavier star of the binary.  In Table~\ref{table:two}, we 
summarize the estimated physical parameters of $M_1$, $M_2$, $M_3$, $\dl$, $a_{\perp,2}$, and $a_{\perp,3}$.
We find that the {\it Gaia} measurement of the vector source proper motion, $\muvec_{\rm S}(l,b)=
(-11.4,+0.7)\,{\rm mas\,yr^{-1}}$, when combined with the scalar lens-source relative proper motion 
from Equation~(\ref{eq7}), strongly constrains the lens to lie in the bulge.  Here $\muvec_{\rm S}(l,b)$ 
denotes the source proper motion vector in the galactic coordinates, and it is related to 
$\muvec_{\rm S}({\rm RA, dec})$ by
\begin{equation}
\left( \begin{array}{c}
\mu_{{\rm S},l}  \\
\mu_{{\rm S},b}  
\end{array} \right) 
=
\left( \begin{array}{cc}
 \cos\eta & \sin\eta \\
-\cos\eta & \cos\eta
\end{array} \right)
\left( \begin{array}{c}
\mu_{\rm S,RA}  \\
\mu_{\rm S,dec}  
\end{array} \right), 
\label{eq11}
\end{equation}
where $\eta\sim 57.6^\circ$ is the tilt angle of the galactic plane with respect to the celestial equator.
This is because the Galactic model contains relatively few disk lenses with $\mu_{\rm L}(l) \lesssim 
-5.8~{\rm mas~yr^{-1}}$. Specifically, we find that when the {\it Gaia} measurement is ignored, the 
probability of a disk lens is 28\%. However, after including the {\it Gaia} measurement, the probability 
of a disk lens is only $\sim 1\%$.

\begin{deluxetable}{lcc}
\tablecaption{Physical lens parameters\label{table:two}}
\tablewidth{240pt}
\tablehead{
\multicolumn{1}{c}{Parameter}     &
\multicolumn{1}{c}{3L1S}          &        
\multicolumn{1}{c}{2L2S}         
}
\startdata                                 
$M_1$ ($M_\odot$)       &   $0.27^{+0.27}_{-0.12}$    &  $0.12^{+0.12}_{-0.05}$   \\
$M_2$ ($M_\odot$)       &   $0.10^{+0.10}_{-0.04}$    &  $0.045^{+0.045}_{-.019}$ \\
$M_3$ ($M_{\rm J}$)     &   $0.51^{+0.51}_{-0.23}$    &  --                       \\
$\dl$ (kpc)             &   $6.98^{+0.91}_{-0.90}$    &  $6.97^{+0.70}_{-0.86}$   \\
$a_{\perp,2}$ (AU)      &   $5.19^{+5.87}_{-4.52}$    &  $3.52^{+3.87}_{-3.09} $ \\
$a_{\perp,3}$ (AU)      &   $1.23^{+1.39}_{-1.07}$    &  --  
\enddata                            
\end{deluxetable}

The Bayesian posterior distributions obtained with the 2L2S solution are also presented in 
Figure~\ref{fig:eleven} (red unshade curves).  The estimated masses of the lens components and the 
distance to the lens system are 
\begin{equation}
M_1=  0.12^{+0.12}_{-0.05}~M_\odot,\ \
M_2=  0.045^{+0.045}_{-.019}~M_\odot,\ \ 
\label{eq12}
\end{equation}
and
\begin{equation}
\dl = 6.97^{+0.70}_{-0.86}~{\rm kpc},
\label{eq13}
\end{equation}
respectively.  Then, the lens is a binary composed of an M dwarf and a brown dwarf located in the bulge.  
The lens mass from the 2L2S solution, $M_{\rm 2L2S}=M_1+M_2$, is smaller than the value estimated from the 
3L1S solution, $M_{\rm 3L1S}$, because the Bayesian input value of the Einstein radius, $\theta_{\rm E, 2L2S}
\sim 0.18$~mas, is smaller than value of the 3L1S solution, $\theta_{\rm E, 3L1S}\sim 0.29$~mas.  Considering 
that the lens distances for the two solutions are similar to each other, it is found that the lens masses 
estimated from the two solutions are approximately in the relation $M_{\rm 2L2S} \sim (\theta_{\rm E, 2L2S}/
\theta_{\rm E, 3L1S} )^2 M_{\rm 3L1S}$.  The physical lens parameters of the 2L2S solution are also 
summarized in Table~\ref{table:two}.

\section{Discussion}\label{sec:six}

If the 3L1S model is correct, OGLE-2019-BLG-0304LAbB has the second lowest formal significance,
$\Delta\chi^2 = 48.3$, of any reported microlensing planet, with the lowest being $\Delta\chi^2 =47$ 
for OGLE-2018-BLG-0677 \citep{Herrera2020} and the previous second-lowest being $\Delta\chi^2 =170$ 
for KMT-2018-BLG-1025 \citep{Han2021}.  Moreover, OGLE-2019-BLG-0304LAbB is a more complex (3L1S) 
system, whereas the other low formal-significance systems were 2L1S.  This distinction is important
because, of the seven previously claimed binary+planet microlensing systems, two were subsequently 
shown to be spurious, namely MACHO-97-BLG-41 \citep{Bennett1999, Albrow2000, Jung2013} and 
OGLE-2013-BLG-0723 \citep{Udalski2015, Han2016}.  In both cases, the original models did not incorporate 
orbital motion.  The additional features in the light curve, that were inconsistent with static 2L1S 
models (and so were attributed to a third body), were found to be explained by the motion of a caustic 
(due to the underlying binary motion) within the context of orbital-motion models.  This was the major 
motivation for our check in Sect.~\ref{sec:three-one} of 2L1S orbital-motion models, which could not 
explain the additional ``planetary'' anomalies. However, it is also important to understand at a deeper 
level why orbital motion is strongly constrained to the point that it cannot explain these anomalies.

Table~\ref{table:three} shows the results of the fits that exclude data from the second bump. For 
the 2L1S model, the parameters $(s_2,\alpha_2) = (3.900\pm 0.040,169^\circ\hskip-2pt .88\pm 0^\circ
\hskip-2pt .44)$ essentially predict the location of the second bump, without any direct light-curve 
information about the bump, which peaks 61 days later. In the full-data (static) model, this bump 
location fixes these parameters with much greater precision: $(s_2,\alpha_2) = (3.784\pm 0.015,169^\circ
\hskip-2pt .25\pm 0^\circ\hskip-2pt .11)$.  We will comment on the $3\,\sigma$ discrepancy in $s_2$ below. 
But for the moment, the main point is that the overall agreement between this bump-free ``prediction'' 
and the actual location of the bump 61 days later, implies that the orbital motion on the $1$~day 
timescales of the main peak (where the planetary anomaly appears) is extremely well constrained.

As shown by the Figure~\ref{fig:two} residuals, the only really pronounced deviations from the 2L1S 
model comes from the three KMTC points on each of HJD$^\prime =$ 8542.xx and 8543.xx.  These contribute 
$\Delta\chi^2=7$ and $\Delta\chi^2=18$ to the total $\Delta\chi^2=48$. Moreover, they could contribute 
indirectly to the $\Delta\chi^2=10$ from the early light curve (because the KMTS data, with similar 
coverage during this period, do not show any significant $\Delta\chi^2$).  We therefore focus particular 
attention on these two nights.

\begin{deluxetable}{lcccc}
\tablecaption{2L1S and 3L1S models without the second bump\label{table:three}}
\tablewidth{240pt}
\tablehead{
\multicolumn{1}{c}{Parameter}                      &
\multicolumn{1}{c}{2L1S}                           &
\multicolumn{1}{c}{3L1S}                           
}
\startdata   
$\chi^2$                        &     $4728.8              $   &  $4699.5             $     \\
$t_0$ (${\rm HJD}^\prime$)      &     $8543.617 \pm 0.005  $   &  $8543.606 \pm 0.006 $     \\
$u_0$ ($10^{-3}$)               &     $1.84 \pm 0.46       $   &  $0.98 \pm 0.55      $     \\
$\te$ (days)                    &     $17.43 \pm 0.15      $   &  $17.62 \pm 0.19     $     \\
$s_2$                           &     $3.900 \pm 0.040     $   &  $3.837 \pm 0.063    $     \\
$q_2$                           &     $0.436 \pm 0.015     $   &  $0.396 \pm 0.022    $     \\ 
$\alpha$ (rad)                  &     $2.965 \pm 0.008     $   &  $2.965 \pm 0.009    $     \\
$s_3$                           &      --                      &  $0.879 \pm 0.008    $     \\
$q_3$ ($10^{-3}$)               &      --                      &  $1.67 \pm 0.26      $     \\
$\psi$ (rad)                    &      --                      &  $2.514 \pm 0.109    $     \\
$\rho$ ($10^{-2}$)              &     $1.97 \pm 0.04       $   &  $1.88 \pm 0.05      $     
\enddata                            
\tablecomments{ 
${\rm HJD}^\prime\equiv {\rm HJD}-2450000$.
\smallskip
}
\end{deluxetable}

We note that the crescent moon passed within $6^\circ\hskip-2pt .1$ of the event at HJD$^\prime =
8542.37$, and during the three observations on 8542.xx, the moon was separated by $8^\circ\hskip-2pt 
.4$--$8^\circ \hskip-2pt .8$, which gave rise to background counts of about 1900 per pixel (compared 
to $\sim 500$ for dark time and $\sim 25,000$ when the full moon is in the bulge).  However, on the 
next night, which yielded a much larger ``planetary signal'', the Moon was 
$19^\circ\hskip-2pt .1$--$19^\circ\hskip-2pt .6$ from the event, and the background was only about 
800.  Hence, it is unlikely that these relatively normal (especially on the second night) observing 
conditions could be responsible for the planetary anomaly.  For double check, we investigate the 
effect of the moon by additionally (1) conducting visual inspection of images, (2) probing the 
dependence of the photometry on the moon phase and distance, and (3) comparing photometric result 
with those of other comparison stars.  From this, we find nothing irregular about them.

Finally, the $\Delta\chi^2=19$ improvement when the bump is included (see Figure~\ref{fig:nine}) lends 
added weight to the reality of the signal.  This improvement arises from the greater consistency between 
the ``predicted'' and actual forms of the late-time bump when the planet is included in the model.
That is, regardless of the origin of the deviations in the KMTC light curve (i.e., whether a planet or 
some random systematics), the 2L1S and 3L1S fits will try to adjust their parameters to accommodate these 
deviations. If the ``planet'' is just the result of accommodating random systematics in the 3L1S fit, 
then the 3L1S prediction should, on average, be no better or worse than the 2L1S prediction. Using the 
parameters presented in Tables~\ref{table:one} and \ref{table:three}, we find that the differences in 
the parameters between the two sets of models obtained with the full data and the partial data without 
the second bump are 
$\Delta (s_2,q_2,\alpha) =(0.116,0.044,0^\circ\hskip-2pt .63)\pm (0.040,0.015,0^\circ\hskip-2pt .44)$ 
for the 2L1S model and 
$\Delta (s_2,q_2,\alpha) =(0.104,0.032,0^\circ\hskip-2pt .29)\pm (0.063,0.022,0^\circ\hskip-2pt .50)$ 
for the 3L1S model.  In all cases, there is improved agreement.  And, by adding the planet, the $\chi^2$ 
drops from 19.1 to 5.3 for three degrees of freedom.  We therefore conclude that the residual from the 
2L1S model cannot be ascribed to the lens orbital motion, although its 3L1S or 2L2S origin cannot be 
firmly distinguished.

\section{Conclusion}\label{sec:seven}

We analyzed the microlensing event OGLE-2019-BLG-0304. The light curve of the event showed two 
distinctive features, in which the main peak appeared to exhibit deviations caused by finite-source 
effects, and there existed a second bump appearing $\sim 61$~days after the main peak. The
fit with a finite-source single-lens model excluding the second bump left substantial residuals 
in the peak region, indicating that a more sophisticated model was needed to explain not only the 
second bump but also the deviations in the peak region.  A 2L1S model could explain the overall 
features of the light curve by significantly reducing the residuals in both regions of the main peak 
and the second bump, but the model still left subtle but noticeable residuals in the peak region. 
We found that the residuals could be explained by the presence of either a planetary companion 
located close to the primary of the lens  or an additional close companion to the source.  Although 
the 3L1S model is favored over the 2L2S model, firm resolution of the degeneracy between the two models 
was difficult with the photometric data.  Therefore, the event well illustrated the need for thorough 
model testing in the interpretation of lensing events with complex features.

We estimated the physical lens parameters expected from the two degenerate solutions by conducting 
Bayesian analysis.  According to the 3L1S interpretation, the lens is a planetary system, in which 
a planet with a mass $0.51^{+0.51}_{-0.23}~M_{\rm J}$ is in an S-type orbit  around a binary 
composed of stars with masses $0.27^{+0.27}_{-0.12}~M_\odot$ and $0.10^{+0.10}_{-0.04}~M_\odot$.  
According to the 2L2S interpretation, the source is composed of G- and K-type giant stars, and the 
lens is composed of a low-mass M dwarf and a brown dwarf with masses $0.12^{+0.12}_{-0.05}~M_\odot$ 
and $0.045^{+0.045}_{-.019}~M_\odot$, respectively.

\acknowledgments
Work by C.H. was supported by the grants of National Research Foundation of Korea
(2019R1A2C2085965 and 2020R1A4A2002885).
Work by A.G. was supported by JPL grant 1500811.
This research has made use of the KMTNet system operated by the Korea
Astronomy and Space Science Institute (KASI) and the data were obtained at
three host sites of CTIO in Chile, SAAO in South Africa, and SSO in
Australia.
The OGLE project has received funding from the National Science Centre, Poland, grant
MAESTRO 2014/14/A/ST9/00121 to AU.


\begin{thebibliography}{}
\bibitem[Alard \& Lupton(1998)]{Alard1998} Alard, C., \& Lupton, R.~H.\ 1998, \apj, 503, 325
\bibitem[Albrow(2017)]{Albrow2017} Albrow, M.\ 2017, MichaelDAlbrow/pyDIA: Initial Release on Github,Version v1.0.0, Zenodo, doi:10.5281/zenodo.268049
\bibitem[Albrow et al.(2000)]{Albrow2000} Albrow, M.~D., Beaulieu, J.-P., Caldwell, J.~A.~R., et al.\ 2000, \apj, 534, 894 
\bibitem[Albrow et al.(2009)]{Albrow2009} Albrow, M., Horne, K., Bramich, D.~M., et al.\ 2009, \mnras, 397, 2099
\bibitem[An(2005)]{An2005} An, J.~H.\ 2005, \mnras, 356, 1409
\bibitem[Assef et al.(2006)]{Assef2006} Assef, R.~J., Gould, A., Afonso, C.\ et al., 2006, \apj, 649, 954
\bibitem[Bennett et al.(2014)]{Bennett2014} Bennett, D.~P., Batista, V., Bond, I.~A., et al.\ 2014, \apj, 785, 155
\bibitem[Bennett et al.(1999)]{Bennett1999} Bennett, D.~P., Rhie, S.~H., Becker, A.~C., et al.\ 1999, Nature, 402, 57 
\bibitem[Bennett et al.(2016)]{Bennett2016} Bennett, D.~P., Rhie, S.~H., Udalski, A., et al.\ 2016, \aj, 152, 125
\bibitem[Bennett et al.(2020)]{Bennett2020} Bennett, D.~P., Udalski, A., Bond, I.~A.\ 2020, \aj, 160, 72
\bibitem[Bensby et al.(2013)]{Bensby2013} Bensby, T., Yee, J.~C., Feltzing, S., et al.\ 2013, \aap, 549, 147
\bibitem[Bessell \& Brett(1988)]{Bessell1988} Bessell, M.~S., \& Brett, J.~M.\ 1988, \pasp, 100, 1134
\bibitem[Bozza(1999)]{Bozza1999} Bozza, V.\ 1999, \aap, 348, 311
\bibitem[Chung et al.(2005)]{Chung2005} Chung, S.-J., Han, C., Park, B.-G., et al.\ 2005, \apj, 630, 535
\bibitem[Correia et al.(2005)]{Correia2005} Correia, A.~C.~M., Udry, S., Mayor, M., Laskar, J., Naef, D., Pepe, F., Queloz, D., \& Santos, N.~C.\ 2005, \aap, 440, 751
\bibitem[Dan{\u e}k \& Heyrovsk\'y(2015)]{Danek2015} Dan{\u e}k, K., \& Heyrovsk\'y, D.\ 2015, \apj, 806, 99
\bibitem[Dan{\u e}k \& Heyrovsk\'y(2019)]{Danek2019} Dan{\u e}k, K., \& Heyrovsk\'y, D.\ 2019, \apj, 880, 72
\bibitem[Dominik(1998)]{Dominik1998} Dominik, M.\ 1998, \aap, 329, 361
\bibitem[Dominik(1999)]{Dominik1999} Dominik, M.\ 1999, \aap, 349, 108
\bibitem[Doyle et al.(2011)]{Doyle2011} Doyle, L.~R., Carter, J.~A., Fabrycky, D.~C., et al.\ 2011, Science, 333, 1602
\bibitem[Gaia Collaboration et al.(2018)]{Gaia2018} Gaia Collaboration, Brown, A.~G.~A., Vallenari, A., et al.\ 2018, \aap, 616, A1
\bibitem[Gaudi et al.(1998)]{Gaudi1998} Gaudi, B.~S., Naber, R.~M., \& Sackett, P.~D.\ 1998, \apj, 502, L33 
\bibitem[Gould(2000)]{Gould2000} Gould, A.\ 2000, \apj, 535, 928
\bibitem[Gould(1992)]{Gould1992} Gould, A.\ 1992, \apj, 392, 442
\bibitem[Gould et al.(2014)]{Gould2014} Gould, A., Udalski, A., Shin, I.-G., et al.\ 2014, Science, 345, 46
\bibitem[Griest \& Safizadeh(1998)]{Griest1998} Griest, K., \& Safizadeh, N.\ 1998, \apj, 500, 37
\bibitem[Han(2008)]{Han2008} Han, C.\ 2008, \apjl, 676, L53
\bibitem[Han et al.(2016)]{Han2016} Han, C., Bennett, D.~P., Udalski, A., \& Jung, Y.~K.\ 2016, \apj, 825, 8 
\bibitem[Han et al.(2001)]{Han2001} Han, C., Chang, H.-Y., An, J. H., \& Chang, K.\ 2001, \mnras, 328, 986
\bibitem[Han \& Gould(1995)]{Han1995} Han, C., \& Gould, A.\ 1995, \apj, 447, 53
\bibitem[Han \& Gould(2003)]{Han2003} Han, C., \& Gould, A.\ 2003, \apj, 592, 172
\bibitem[Han et al.(2020a)]{Han2020a} Han, C., Lee, C.-U., Udalski, A., et al.\ 2020a, \aj, 159, 48
\bibitem[Han et al.(2013)]{Han2013} Han, C., Udalski, A., Choi, J.-Y., et al.\ 2013, \apjl, 762 , L28
\bibitem[Han et al.(2017)]{Han2017} Han, C., Udalski, A., Gould, A., et al.\ 2017, \aj, 154, 223
\bibitem[Han et al.(2021)]{Han2021} Han, C., Udalski, A., Lee, C.-U., et al.\ 2021, \aap, 649, A90
\bibitem[Hwang et al.(2013)]{Hwang2013} Hwang, K.-H., Choi, J.-Y., Bond, I.~A., et al.\ 2013, \apj, 778, 55
\bibitem[Herrera-Mart\'in et al.(2020)]{Herrera2020} Herrera-Mart\'in, A., Albrow, M.~D., Udalski, A., et al.\ 2020, \aj, 159, 256
\bibitem[Jung et al.(2013)]{Jung2013} Jung, Y.~K., Han, C., Gould, A., \& Maoz, D.\ 2013, \apjl, 768, L7 
\bibitem[Kervella et al.(2004)]{Kervella2004} Kervella, P., Th\'evenin, F., Di Folco, E., \& S\'egransan, D.\ 2004, \aap, 426, 29
\bibitem[Kim et al.(2016)]{Kim2016} Kim, S.-L., Lee, C.-U., Park, B.-G., et al.\ 2016, JKAS, 49, 37
\bibitem[Kim et al.(2018)]{Kim2018} Kim, D.-J., Kim, H.-W., Hwang, K.-H., et al.\ 2018, \aj, 155, 76                          
\bibitem[Lee et al.(2008)]{Lee2008} Lee, D.-W., Lee, C.-U., Park, B.-G., Chung, S.-J., Kim, Y.-S., Kim, H.-I., \& Han, C.\ 2008, \apj, 672, 623
\bibitem[Lee et al.(2009)]{Lee2009} Lee, J.~W., Kim, S.-L., Kim, C.-H., Koch, R.~H., Lee, C.-U., Kim, H.-I., \& Park, J.-H.\ 2009, \aj, 137, 3181
\bibitem[Morris(1985)]{Morris1985} Morris, S.~L.\ 1985, \apj, 295, 143
\bibitem[Nataf et al.(2013)]{Nataf2013} Nataf, D.~M., Gould, A., Fouqu\'e, P., et al.\ 2013, \apj, 769, 88
\bibitem[Poleski et al.(2014)]{Poleski2014} Poleski, R., Skowron, J., Udalski, A., et al.\ 2014, \apj, 795, 42
\bibitem[Rhie(2002)]{Rhie2002} Rhie, S.~H.\ 2002, arXiv:astro-ph/0202294
\bibitem[Di Stefano \& Scalzo(1999)]{Stefano1999} Di Stefano, R., \& Scalzo, R.~A.\ 1999, \apj, 512, 579
\bibitem[Thorsett et al.(1993)]{Thorsett1993} Thorsett, S.~E., Arzoumanian, Z., \& Taylor, J.~H.\ 1993, \apjl, 412, L33
\bibitem[Tomaney \& Crotts(1996)]{Tomaney1996} Tomaney, A.~B., \& Crotts, A.~P.~S.\ 1996, \aj, 112, 2872
\bibitem[Udalski(2003)]{Udalski2003} Udalski, A.\ 2003, Acta Astron., 53, 291
\bibitem[Udalski et al.(2015)]{Udalski2015} Udalski, A., Jung, Y.~K., Han, C., et al.\ 2015, \apj, 812, 47 
\bibitem[Udalski et al.(1994)]{Udalski1994} Udalski, A., Kubiak, M., Szyma\'nski, M., Ka{\l}u\.zny, J., Mateo, M., Krzemi\'nski, W.\ 1994, Acta Astron., 44, 317
\bibitem[Wo\'zniak(2000)]{Wozniak2000} Wo\'zniak, P. R. 2000, Acta Astron., 50, 42
\bibitem[Yee et al.(2012)]{Yee2012} Yee, J. C., Shvartzvald, Y., Gal-Yam, A., et al.\ 2012, \apj, 755, 102
\bibitem[Yoo et al.(2004)]{Yoo2004} Yoo, J., DePoy, D.~L., Gal-Yam, A., et al.\ 2004, \apj, 603, 139
\bibitem[Zhang et al.(2019)]{Zhang2020} Zhang, X., Zang, W., Udalski, A., et al.\ 2020, \aj, 159, 116 


\end{thebibliography}
\end{document}